# A Mixed-Signal Structured AdEx Neuron for Accelerated Neuromorphic Cores

Syed Ahmed Aamir, *Student Member, IEEE*, Paul Müller, Gerd Kiene, Laura Kriener, Yannik Stradmann, Andreas Grübl, Johannes Schemmel, *Member, IEEE*, and Karlheinz Meier, *Member, IEEE*

*Abstract*—Here we describe a multi-compartment neuron circuit based on the Adaptive-Exponential I&F (AdEx) model, developed for the second-generation BrainScaleS hardware. Based on an existing modular Leaky Integrate and Fire (LIF) architecture designed in 65 nm CMOS, the circuit features exponential spike generation, neuronal adaptation, inter-compartmental connections as well as a conductance-based reset. The design reproduces a diverse set of firing patterns observed in cortical pyramidal neurons. Further, it enables the emulation of sodium and calcium spikes, as well as N-Methyl-D-Aspartate (NMDA) plateau potentials known from apical and thin dendrites. We characterize the AdEx circuit extensions and exemplify how the interplay between passive and non-linear active signal processing enhances the computational capabilities of single (but structured) on-chip neurons.

## I. Introduction

Low-complexity point neuron models are mathematically tractable and lead to a cost-effective in-silico implementation [1]. In order to realize large-scale neuromorphic systems, they offer efficient utilization of silicon die area and maintain an energy-efficient solution. The classical LIF neuron [2], [3] has therefore been the popular choice in many first-generation neuromorphic systems [4]–[6]. However, compared to experimental spike data, the LIF model falls short of predicting the spike-times and reproduce the diverse properties of biological neurons. Extended threshold-based models, such as two-dimensional models add a second slow variable or additional state variable to define the membrane evolution. When coupled with a non-linear term [7]–[10], they can more accurately predict the physiological data [11]–[13].

The role of single neurons in computation is however more sophisticated than linear pre-processing of synaptic inputs or their integration on the membrane that leads to an output event once a threshold is reached [14], [15]. More specifically, passive dendrites [16] provide computational functions such as independent computation of local non-linear operations [17], directional selectivity in retinal ganglion cells [18], [19] or coincidence detection in auditory pathway [20], [21] to state a few. In addition, the active properties of dendrites contribute to an internal feedback mechanism with back-propagating action potentials [22]–[24] that induce Long Term Potentiation (LTP) in CA1 [25], Layer 5 neurons [26] or evoke broad $Ca^{2+}$ spikes in apical dendrites [27]. Further, in the thin dendrites they contribute to local NMDA plateau potentials [28], [29] marked by an extraordinary duration and significant amplitude, and proposed to play a significant role in cortical information processing and memory consolidation [30].

In this paper, we present a 65 nm CMOS AdEx neuron circuit with multi-compartment emulation capturing properties of passive and non-linear active dendrites. We enhance our modular LIF point neuron architecture characterized in detail in [31], [32] with AdEx extensions [33]. Further, as detailed in [34], in addition to the sodium spikes, the dendritic implementation allows to emulate calcium spikes and NMDA plateau potentials. The neuron description in [33], [34] is based on circuit simulations, while here we demonstrate the measured results from the prototype chip.

The implemented neuron model is designed for integration in the second-generation BrainScaleS 65 nm physical model platform [35], operated ("accelerated" to) 1000 times faster than biological real-time. The presented analog continuous-time neuron in this work is measured on a 65 nm prototype chip that implements a scaled-down array of 32 neurons, connected to 32 × 32 synapses. A finalized version of this circuit will feature in the larger HICANN-DLS chip – the basic building block of the second-generation BrainScaleS hardware targeting a wafer-scale implementation [36]. We omit here the description and architecture of the prototype chip [37] and detailed characterization of the LIF neuron (see [31], [32] for details). In the subsequent sections, we describe the hardware neuron model and the corresponding implemented architecture (Sec. II and Sec. III). The circuit description of the AdEx and multi-compartment extensions are detailed in Sec. IV, followed by experimental results in Sec. V. A discussion on the presented neuron and its comparison with other architectures concludes the paper (Sec. VI and Sec. VII).

## II. Neuron Model

The neuron circuit adheres to the AdEx point-neuron model. However it replaces a fixed reset with a conductance-based reset. The evolution of the neuron membrane in the AdEx

Funded by the European Union Seventh Framework Programme ([FP7/2007-2013]) under grant agreement nos. 604102 (HBP), 269921 (BrainScaleS), the Horizon 2020 Framework Programme ([H2020/2014-2020]) under grant agreement no. 720270 (HBP) as well as from the Manfred Stärk Foundation.

All authors are with the Kirchhoff Institute for Physics, Heidelberg University, D-69120 Heidelberg, Germany (email: aamir@kip.uni-heidelberg.de).

model is described by a two-variable equation [9] given by

$$C_{\text{mem}}\frac{dV_{\text{mem}}}{dt} = I - w - g_{\text{leak}}(V_{\text{mem}} - V_{\text{leak}})$$
$$+ g_{\text{leak}}\Delta_T \exp\left(\frac{V_{\text{mem}} - V_T}{\Delta_T}\right) \quad (1)$$

$$\tau_{\text{w}}\frac{dw}{dt} = a(V_{\text{mem}} - V_{\text{leak}}) - w \quad (2)$$

where $I$ is the sum of external, synaptic and inter-compartment input currents, $w$ is the adaptation current. The last two terms in Eq. 1 model the leak and exponential currents. $C_{\text{mem}}$ is the membrane capacitance, $V_{\text{mem}}$ is the membrane potential, $g_{\text{leak}}$ and $V_{\text{leak}}$ are the leak conductance and leak potential respectively, $a$ denotes the subthreshold adaptation, $V_T$ is the exponential threshold and $\Delta_T$ is its slope factor. At spike time, the membrane is reset to a specified reset potential, and the adaptation variable $w$ is updated by a current $b$, such that $w \rightarrow w + b$.

For hardware realization, a transformation is performed for the adaptation term [38] equating the adaptation output current,

$$w = a(V_{\text{w}} - V_{\text{leak}}) \quad (3)$$

whose substitution modifies Eq. 2 as

$$-C_{\text{w}}\frac{dV_{\text{w}}}{dt} = g_{\text{w}}(V_{\text{w}} - V_{\text{mem}}) \quad (4)$$

The dendritic structure is implemented using controlled conductance between adjacent isopotential membranes. The inter-compartmental current flowing through the tunable conductance connecting the two membranes is expressed as:

$$I_{\text{ic}} = g_{\text{ic}}(V_{\text{i}} - V_{\text{j}}) \quad (5)$$

where $g_{\text{ic}}$ is the inter-compartmental conductance between two shunted membranes with potentials $V_{\text{i}}$ and $V_{\text{j}}$.

## III. NEURON ARCHITECTURE

### A. Multi-Compartment Architecture

The BrainScaleS neuron can merge its dendritic tree by connecting adjacent neuron compartments in the array – resulting in a larger membrane capacitor with increased inputs. Additionally, this AdEx point-neuron emulation can be extended to multiple compartments in the circuit using an inter-compartmental resistor. This architecture is highlighted in Fig. 1. It shows a matrix of synapse circuits connected to an array of neurons. Every neuron circuit is shown receiving excitatory and inhibitory input lines from a single column of synapses in the matrix. Inside the neuron array, a single column corresponds to a point neuron. The merged dendritic tree is realized by using the switches $S_{\text{ic}}$, whereas multiple compartments may be realized by utilizing inter-compartmental resistors $R_{\text{ic}}$. Each neuron compartment may be configured to emulate either Sodium spikes or broad plateau potentials [34].

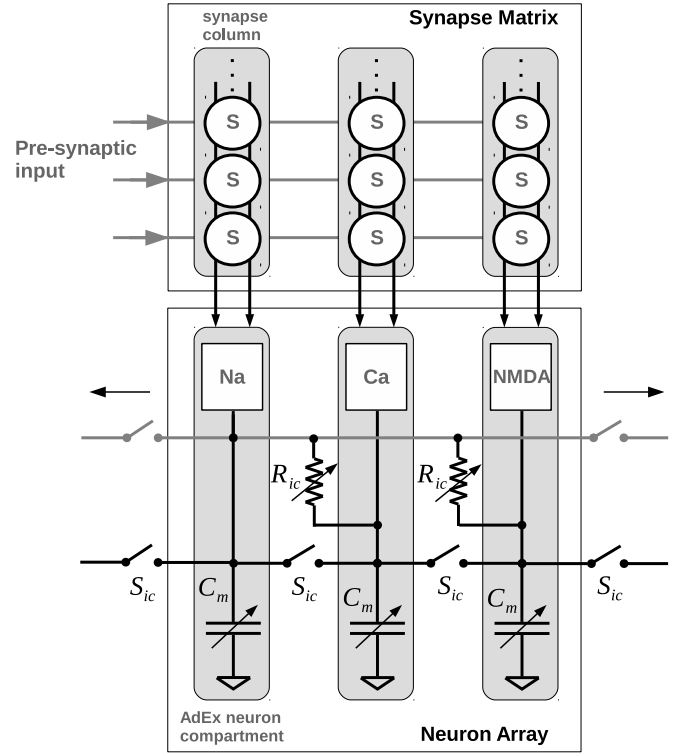

Fig. 1: A conceptual view of the multi-compartment, high-input count AdEx neuron array architecture connected to a matrix of input synapses that receive asynchronous pre-synaptic inputs – adapted from [34].

### B. Point Neuron Architecture

The LIF neuron circuit proposed and characterized in [31], [32] introduces a modular architecture, where individual subcircuits are interconnected using transmission-gate switches. The multi-compartment, AdEx neuron with non-linear dendrites is an extended enhancement of this base LIF architecture.

The architecture of a single integrated neuron circuit connected to a synapse column is shown in Fig. 2a. On the left half, the figure shows a single synapse column relaying input current on two wires – one for excitatory inputs and another for inhibitory inputs. Each asynchronous pre-synaptic network event enables a 6-bit DAC for a 4 ns duration. The 6-bit DAC code is the synaptic weight that modulates the amplitude of the input pulse event.

The right half shows the AdEx neuron circuit which receives the two input lines from each synapse column and implements synaptic dynamics within the synaptic input subcircuits. The current output of the synaptic input circuits (excitatory/inhibitory) is integrated on the neuronal membrane, formed by a 6-bit tunable capacitor labeled $C_{\text{mem}}$. The neuron introduces a membrane leak circuit using a source-degenerated OTA in unity gain feedback with a conductance $g_{\text{m}}$ that models $g_{\text{leak}}$. During the refractory period, the circuit utilizes this very leak conductance as reset conductance by enabling higher conductance (approx. $\times 10\, g_\text{m}$). When the membrane potential $V_\text{mem}$ on the capacitor $C_\text{mem}$ reaches a specified threshold $V_\text{thresh}$, the comparator toggles its output to logic high ($V_\text{OH}$). The output of the comparator triggers a pulse with a variable pulse-width duration that is fed to the merged leak/reset circuit. This pulse input connects the membrane to the reset potential $V_\text{rst}$ via the OTA conductance ($g_\text{m}$), as well as enabling its high-conductance mode [32] (for $\times 10 g_\text{m}$ reset conductance). As soon as the membrane ($V_\text{mem}$) resets, the comparator outputs logic low ($V_\text{OL}$). At the end of refractory period, the OTA switches back to low-conductance mode where the leak conductance is $g_\text{m}$. This circuit realizes the conductance-based reset as opposed to a fixed reset suggested by the I&F neurons to model the relative conductances between different compartments – which, for example, enables the effect of synaptic excitation and inhibition on the neuron compartment during plateau potentials. If we neglect the synaptic inputs, the height of the plateau potentials is then determined by the potential $V_\text{NMDA}$.

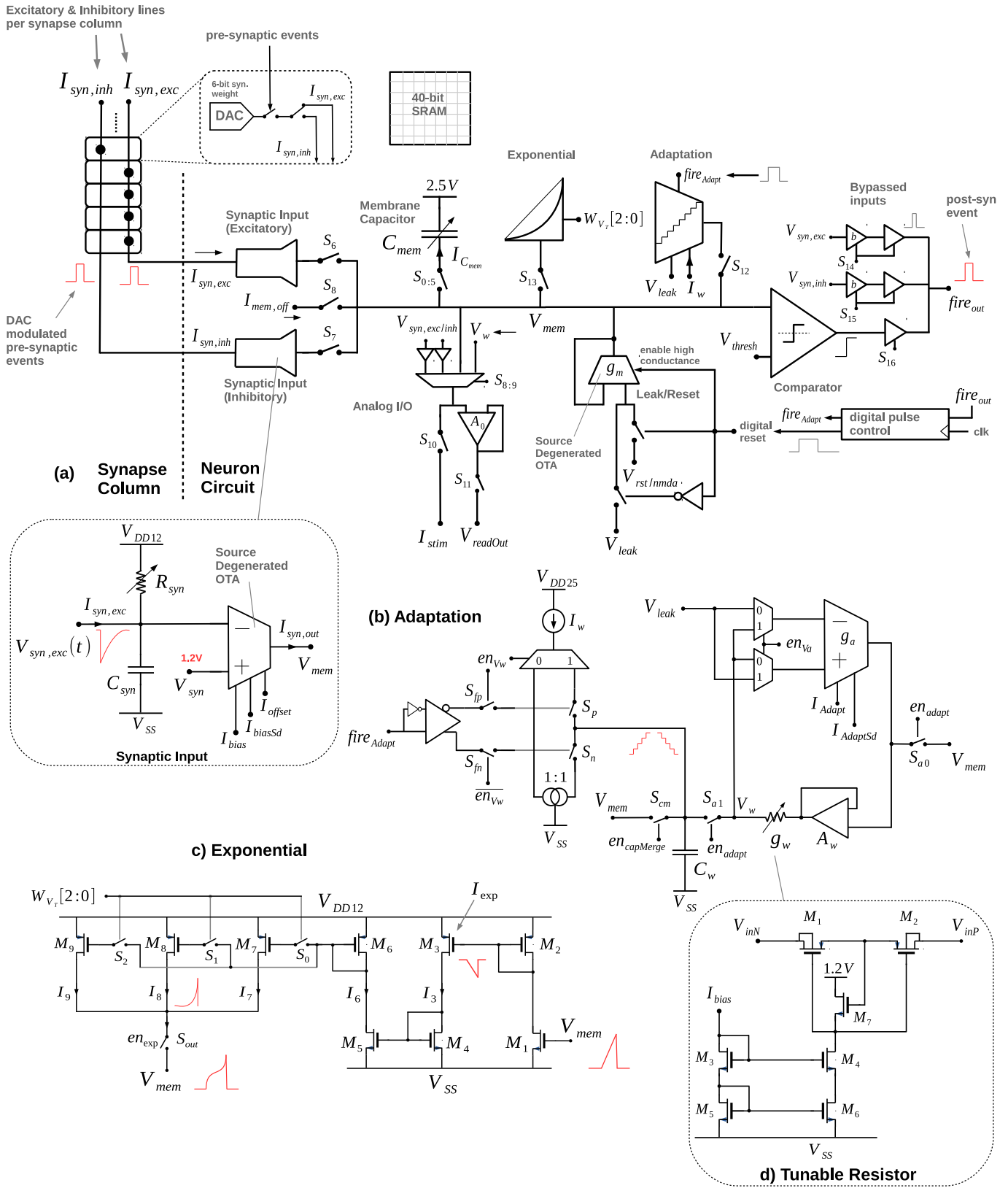

Fig. 2: Simplified schematic diagrams of: a) a single integrated AdEx neuron circuit in a point-neuron configuration receiving input events from a synapse column. b) adaptation circuit. c) exponential circuit. d) tunable resistor used inside the adaptation circuit.

The digital output event also enables the adaptation circuit (for a programmable duration, see fire$_\text{adapt}$) connected to the membrane via the switch $S_{12}$. An exponential circuit integrates a positive exponentially increasing current onto the membrane enabled by the digital implementation of exponential threshold ($W_{V_\text{T}}$), and connected to the membrane via the switch $S_{13}$. The input synaptic events may also bypass membrane integration, evoking a single output event for every single input. These inputs are enabled via the switch $S_{14,15}$ for excitatory and inhibitory *bypassed* inputs. An integrated buffer $A_0$ reads out multiplexed neuronal voltages. These include the membrane potential ($V_\text{mem}$), the voltage at the two synaptic input lines ($V_\text{syn,exc}$, $V_\text{syn,inh}$), as well as the adaptation voltage $V_\text{w}$ inside the adaptation circuit. Similarly, another pin marked $I_\text{stim}$ helps in measurement, for example for measuring output current generated by each subcircuit. The membrane also receives a current input from on-chip current source via $S_8$ (tunable per neuron), to tune the offset currents as well as to stimulate the membrane without an external source.

The neuron's voltage and current bias parameters are stored on-chip in analog memory [39] adjacent to every neuron circuit. A total of 20 dedicated biases tune every neuron (14 currents, 6 voltages) whereas 3 current biases are shared among all neurons in the array. These biases can be tuned with a 10-bit resolution. All digital configuration bits, e.g., the enables of switches are stored in a 40-bit SRAM memory in each neuron. In the implemented chip all SRAM bits are jointly controlled by a global synchronous controller.

### C. Neuron Configurations

To summarize the neuron implements the following firing configurations:

*1) Bypass:* This mode generates an output event for every input synaptic event. This is achieved in a digital I/O configuration and all analog circuits except for the bypass inputs (controlled by $S_{14,15}$) are disabled. Input synaptic events with a large pulse amplitude and pulse-width $> 30$ ns trigger this bypass link. The bypass circuit is an inverter with shifted trip-points, driven by the voltage drop created by the synaptic event at the input synaptic line.

*2) LIF:* This reduces the neuron circuit to the LIF configuration [31] where all non-LIF extensions are disabled. The output tri-state buffer formed by switch $S_{16}$ is enabled.

*3) AdEx:* The standard LIF neuron when supported with AdEx extensions evoke exponential spikes and accelerating and decelerating spike-triggered adaptation. The switches $S_{12,13}$ enable the two terms.

*4) Plateau Potentials:* Along with Sodium spikes emulated by the standard LIF and AdEx configurations [40], the neuron is designed to evoke broad spikes, such as the Calcium spikes or NMDA plateau potentials [34]. When enabled, the membrane is held on to the NMDA plateau potential $V_\text{NMDA}$ for a duration $\tau_\text{NMDA}$, followed by a return to the leak potential $V_\text{leak}$ at the end of pulse duration. Due to a counter-based digital implementation a very large pulse-width duration of up to 1 sec is possible [41].

*5) Multi-compartment & High Input Count:* This configuration utilizes multiple point-neuron circuits in the neuron array [34]. For high input count it shunts multiple adjacent membrane capacitors to realize neurons with increased synaptic input count in the merged neuronal membrane. For multi-compartment realization [34], [42], two independent compartments are connected via an inter-compartmental resistance, for example, to model somatic and dendritic compartments.

## IV. CIRCUIT IMPLEMENTATION

### A. Adaptation Circuit

The adaptation circuit implements accelerating and decelerating spike-triggered adaptation as well as adaptation current given by Eq. 3 and Eq. 4. A simplified circuit schematic is shown in Fig. 2b. The circuit has been inspired from our first-generation design presented in [38].

The top right part of the circuit implements Eq. 3, where the output current $w$ generated by the OTA with conductance $g_\text{a}$ emulates the model's subthreshold conductance parameter $a$. It senses the difference between $V_\text{leak}$ and $V_\text{w}$ at its inputs and switches them with a configuration bit en$_\text{Va}$ to realize negative $g_\text{a}$. The OTA is a source-degenerated architecture identical to the one used in the leak term. (for details, see [32]).

The lower right part of the schematic implements Eq. 4 where the input current $g_\text{w}(V_\text{w} - V_\text{mem})$ is integrated on the adaptation capacitor $C_\text{w}$. The membrane is buffered by an OTA labeled $A_\text{w}$ and a tunable floating conductance $g_\text{w}$ connects this buffered membrane to the capacitor $C_\text{w}$. The circuit is enabled by asserting the bit en$_\text{adapt}$ and dis-asserting en$_\text{capMerge}$. This connects the output to the membrane $V_\text{mem}$ via switch $S_{a0}$ while connecting the capacitor $C_\text{w}$ to the node $V_\text{w}$. The configuration bit en$_\text{capMerge}$ should always stay low during adaptation usage – however when it is disabled (e.g., in LIF mode), the unused adaptation capacitor can be merged with the membrane capacitor $C_\text{mem}$ by enabling en$_\text{capMerge}$ essentially increasing the maximum $C_\text{mem}$ from 2.36 pF to 4.36 pF. The voltage on the capacitor $V_\text{w}$ emulates the adaptation variable in the model. The presence of a tunable conductance implements adaptation time constant $\tau_\text{w} = R_\text{w} C_\text{w}$, where $R_\text{w} = 1/g_\text{w}$.

The left side of the circuit shows a charge pump, where an on-chip bias current $I_w$ is sourced or sinked from the node $V_w$ via the pass transistors $S_p$ and $S_n$ (assuming en$_{adapt}$ = 1). The circuit is triggered by the input event fire$_{adapt}$, whose presence indicates a digital spike event. The pulse-width of fire$_{adapt}$ is variable and an equivalent charge $q = I_w \cdot t_{pulse}$ is integrated (or removed) with every input event from the capacitor $C_w$. The configuration bit en$_{V_w}$ controls whether to source or sink current $I_w$, essentially implementing either decelerating or accelerating output spiking response. The charge pump therefore models the spike-triggered adaptation in terms of integrated voltage, since every output event updates $V_w \rightarrow V_w \pm \Delta V_w$.

The switches $S_{a0}$, $S_{a1}$ and $S_{cm}$ are transmission gate switches. All multiplexed inputs are implemented using transmission-gate multiplexers. The tunable conductance $g_w$ is implemented as a high value tunable floating resistor.

*1) Tunable Floating Resistor:* The adaptation circuit requires a high value floating tunable resistor that implements a resistive range from 16 MΩ to 600 MΩ [33], [43]. The circuit to implement this large resistive range is designed using bulk-drain connected PMOS devices [44], [45]. The simplified schematic of the implemented resistor is shown in Fig. 2d. Between the positive and negative resistor terminals are two series PMOS bulk-drain connected transistors $M_1$ and $M_2$. Each of the two devices connect the bulk terminals to its drain (instead of source) and biased in subthreshold region to contribute a very large resistance. Both of the devices are connected back-to-back such that their source terminals are connected. In this arrangement, only one of the devices is connected as a bulk-drain connected device at a time. When $V_{inN} > V_{inP}$, the device $M_2$ is in bulk-drain configuration, while $M_1$ is in the nominal bulk-source connected configuration and therefore $R_{SD2} \gg R_{SD1}$. This is because denominations of source and drain are with respect to MOS terminal potentials. For $V_{inP} > V_{inN}$, the roles flip and $R_{SD1} \gg R_{SD2}$. A single bias current $I_{bias}$ tunes the resistance. A cascode current mirror realized using the transistors $M_{3-6}$ mirrors the bias current $I_{bias}$, while the transistor $M_7$ sets the bias point.

### B. Exponential Circuit

This circuit integrates an exponentially rising positive current on the membrane capacitor, as governed by the AdEx model. The schematic is shown in Fig. 2c. The circuit uses transistor $M_1$ to sense the membrane potential at its gate. Together with transistor $M_2$, it provides an inverted membrane voltage to $M_3$. It further ensures that this inverted voltage range of membrane potential (from 0.6 V to 1.05 V) biases $M_3$ in its subthreshold region. To maximize this range $M_3$ is chosen as a high-$V_{th}$ device. In subthreshold region the drain current of a transistor for drain-source drop $V_{DS} > 4U_T$ is:

$$I_D = I_0 e^{\frac{V_{GS} - V_{th}}{nU_T}} \quad (6)$$

where $I_0 = 2n\mu C_{ox} \frac{W}{L} U_T^2$ [46]. The parameter $n$ is equivalent to $\frac{C_{ox} + C_{j0}}{C_{ox}}$ where $C_{j0}$ is the junction-depletion capacitance per unit area of a reversed bias diode (0 V bias).

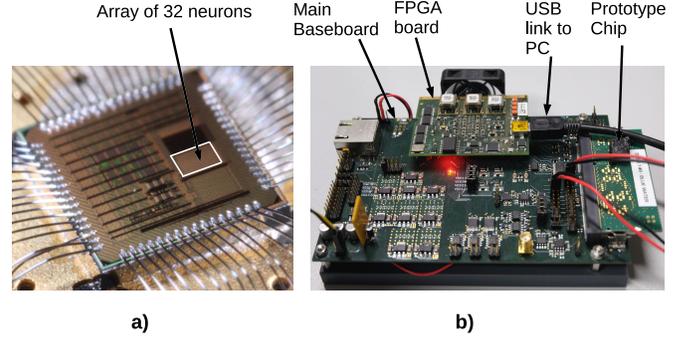

Fig. 3: a) Chip micrograph. b) The prototype evaluation system.

The output drain current of transistor $M_3$ therefore increases exponentially with a linearly rising membrane potential. It is copied by a current mirror formed by $M_{4,5}$ and further by three separate current mirrors formed by $M_{6,7}$, $M_{6,8}$ and $M_{6,9}$. The input/output current ratio of each of the three mirrors is 1:0.25, 1:0.5 and 1:1. The output of all three mirrors is merged into a single output current and integrated onto the membrane. The three mirrors are enabled by a 3-bit digital bus labeled $W_{V_T}$ – essentially realizing the digital exponential threshold of the model parameter $V_T$. The exponential circuit is enabled when the output switch is asserted by bit en$_{exp}$. The circuit limits the output current above a membrane voltage of approx. 1.1 V, since $V_{SD}$ of transistors $M_{7,8,9}$ is below $V_{SD,sat}$ in this case. The 3-bit $W_{V_T}$ as well as all configuration bits are stored in the local SRAM integrated per neuron.

### C. Analog Read-Out

The output of read-out buffers within each neuron circuit (see $V_{readOut}$ in Fig. 2a) are connected via two output buffers (not shown) to the chip pads. Out of the 32 neuron read-out buffers, all odd ones are connected to the first output buffer, while all even read-out buffers are connected to the second output buffer. This scheme enables us to read out any two simultaneous voltages from the neuron array.

## V. EXPERIMENTAL RESULTS

The array of 32 AdEx neurons has been integrated on a prototype chip occupying a total of 286 μm × 376 μm. Each neuron occupies 11.76 × 286 μm of silicon area. The prototype chip is fabricated in a low-K 1P9M 65-nm low-power digital CMOS process with a total die area of 1.9 × 1.9 mm$^2$. The bonded die on the chip daughterboard as well the measurement setup is shown in Fig. 3. The data has been acquired using a Keithley 2635B Sourcemeter for current measurements as well as a LeCroy Wavesurfer 44Xs digital oscilloscope for transient signals. The PCB baseboard designed for measurement comprises of an on-board Xilinx Spartan-6 FPGA that takes command packets via a USB interface. The chip and all neuron parameters are directly programmable via a C++/Python based software setup.

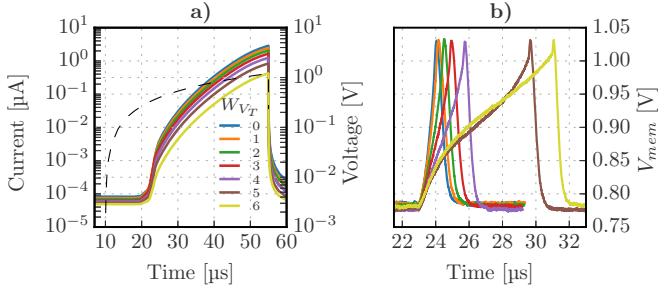

Fig. 4: a) A simulation where the exponential circuit is exposed to a ramp voltage input and its output current is plotted on a logarithmic-scale, while sweeping the exponential threshold parameter $W_{V_T}$. b) Measured successive membrane traces from the prototype chip with swept $W_{V_T}$ [40].

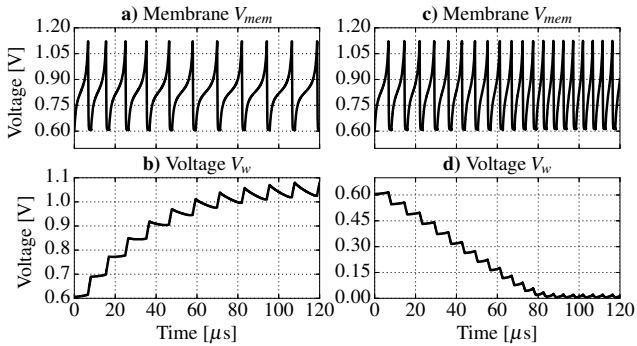

Fig. 5: Spike-triggered adaptation: (a,b) Decelerating membrane as a result of increasing adaptation variable $V_w$. (c,d) Accelerating membrane as a result of decreasing adaptation variable $V_w$.

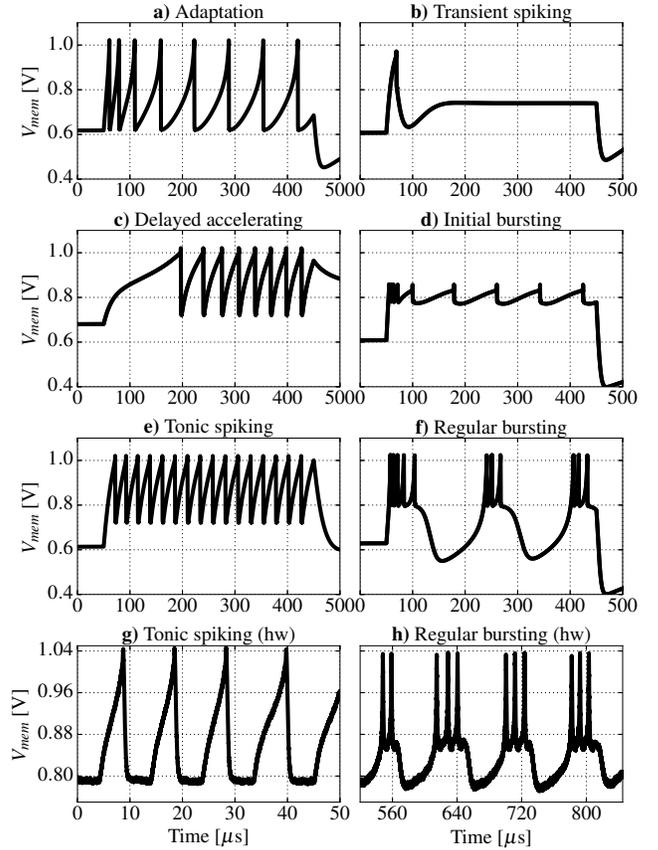

Fig. 6: Firing patterns of the designed AdEx neuron. From a) to f) are those simulated on a circuit netlist [33], [47], whereas g) and h) are measured results from the prototype chip [40].

### A. Exponential

To demonstrate the functionality of the exponential circuit, it is subjected to a slow rising ramp, while its $W_{V_T}$ is swept. Fig. 4a shows this rising voltage ramp (dashed line) in log-scale as well as the corresponding output currents of the exponential circuit. Note that the output exponential current increases as $W_{V_T}$ decreases, with 0 indicating all mirrors enabled. The output log current increases near-linearly for the swept values of $W_{V_T}$ for the required current range from 10 nA to $\geq 1$ µA.

Fig. 4b shows the experimental results of the neuron firing tonically as a results of input current, where the exponential current pulls the membrane voltage up. The digital parameter $W_{V_T}$ is being swept again during successive runs. Note how the strength of exponential current decreases with every decreasing value of $W_{V_T}$. Further, one may also notice the finite time constant that exists during the membrane reset – this is due to the conductance based reset realization.

### B. Adaptation

The accelerating and decelerating spike-triggered adaptation realized by the integration of adaptation circuit is shown in Fig. 5. The adaptation voltage $V_w$ grows from 0.6 V and increments approx 0.1 V with every spike evoked – resulting in decelerating adaptation (Fig. 5a and Fig. 5b). With the same parameters and toggled en$_{V_w}$, a decreasing $V_w$ results which accelerates the membrane (Fig. 5c and Fig. 5d).

The time constant of the adaptation circuit ($\tau_w = R_w C_w$) is tuned with the resistor's tunable bias since the capacitor is fixed at 2 pF. The achievable time constants in the array of 32 neurons is shown in Fig. 7. The short time constants below 100 µs are realized by setting the maximum value of resistor's tunable current bias (1 µA), whereas the long time constants are realized with a minimum available bias of 15 nA. The measured decaying traces show the mismatch between individual curves. This mismatch is more pronounced when the circuit is biased with very small current compared to short time constants where bias current is set much higher. A distribution of minimum and maximum adaptation time constants measured from 31 adaptation circuits on the prototype chip is shown on the right side of Fig. 7.

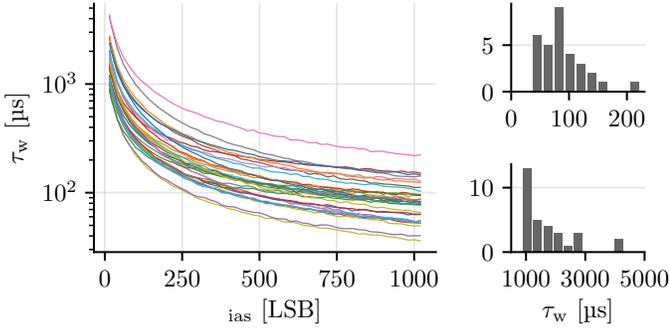

Fig. 7: The adaptation time constant ($\tau_\mathrm{w}$) as a function of bias current of the floating resistor. An uncalibrated distribution of minimum and maximum $\tau_\mathrm{w}$ is plotted for 31 samples.

*C. Example Firing Patterns*

Being a two-variable neuron model, the AdEx circuit reproduces a diverse set of firing patterns [48] known from biological neurons [49], and typically characterized by the response to a step current stimulus. Fig. 6 shows a set of example firing patterns from the designed circuit. Fig. 6a-f show the adaptation, transient spiking, delayed accelerating, initial bursting, tonic spiking as well as regular bursting. These have been simulated on the circuit netlist prior to fabrication on the chip. Fig. 6g,h shows tonic spiking and regular bursting as measured on the prototype chip.

*D. Multi-Compartment Configuration Using Passive Dendrites*

To demonstrate the inter-compartment functionality we configure a chain of resistively connected neurons as highlighted in Fig. 8a. Fig. 9 shows the attenuation of a stimulus—applied to compartment 0—as it passes the neighboring compartments. Tuning the strength of the variable resistor changes the propagation length of the initial stimulus (panels d–f). The varying levels of shift are caused by input offsets of the source-degenerated leak OTA and the readout amplifier in different compartments.

*E. Multi-Compartment Configuration Using Active Dendrites*

To showcase the successful operation of multiple AdEx circuits in a multi-compartment configuration we implement a benchmark that is derived from measurements described in [50]: The authors show that single neurons are capable of acting as coincidence detectors that are dependent on the site of stimulation. In our benchmark, we reduce the biological reference to the following target behavior: A neuron contains three compartments, soma, proximal and distal dendrite. The soma compartment is configured for sodium-like spikes—the exponential term is enabled and the reset potential is below the resting potential, with a comparatively short refractory period. The other compartments are configured as active compartments which can generate plateau potentials, so the refractory period is set to a higher value and the reset potential is set significantly higher than the leakage potential. The resulting array configuration is shown in Fig. 8b. The functional requirement on the neuron is that a somatic stimulus of a certain strength evokes exactly one spike at the soma and no spike at the distal dendrite. Stimulus at the distal dendrite should cause no spikes. Both stimuli combined should evoke a plateau potential in the dendrite and a burst in the soma compartment.

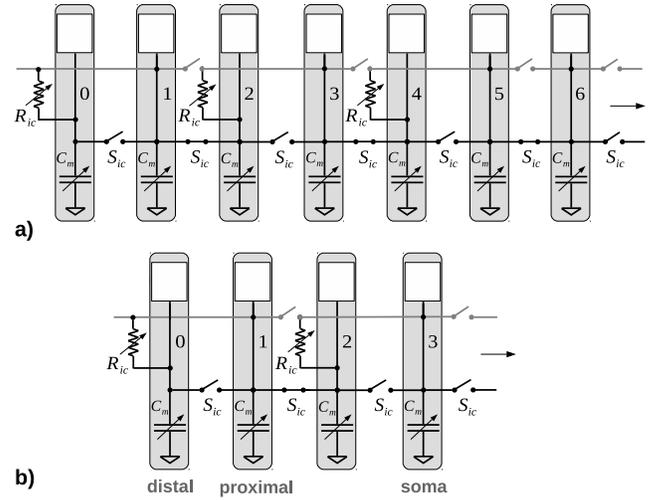

Fig. 8: a) Compartment connectivity for the experiment shown in Fig. 8. Compartment 0 is the start of a chain of compartments, two of which (1 and 2, 3 and 4, etc.) are interconnected directly. Each such pair is connected to its neighbors using the inter-compartment resistors. b) Compartment connectivity for the experiment shown in Fig. 9. Four physical compartments form a chain of three resistively connected parts. The distal (compartment 0) and soma (compartment 3) units receive external synaptic input. Compartments 0 and 1 are configured to produce plateau potentials. Compartment 2 is passive and has its leak and reset circuit disabled. Compartment 3 is configured for Na-spikes.

To achieve this, the refractory times are configured such that the duration of a plateau potential is approximately three times the refractory period in the soma. A coarse parameter sweep is used to determine the approximate parameter values that lead to the desired behavior. Fig. 10 b) shows the result of ten repeated parameter sweeps over the somatic and dendritic firing threshold. The panel indicates the number of experiment runs in which the target behavior is achieved. Although the available region is narrow, the stability of the system for repeated runs is sufficient to reliably reproduce the behavior. Parameters from the center of the stable region are selected and the membrane voltage is recorded for 60 repeated trials. The results are shown in Fig. 10 a): Only one out of 60 samples deviates from the standard response shape (shown with gray color). The height of the voltage spikes in the bottom right panel in Fig. 10 a) differs because the soma compartment is configured with small (one out of 15 clock cycles) holdoff-time at the end of the refractory period (during hold-off the neuron is set to the leak conductance but no new spike may trigger). Consequently, in the presence of a plateau potential in the neighboring compartment the membrane voltage is pulled up above the firing threshold before a new spike is emitted and a reset is triggered.

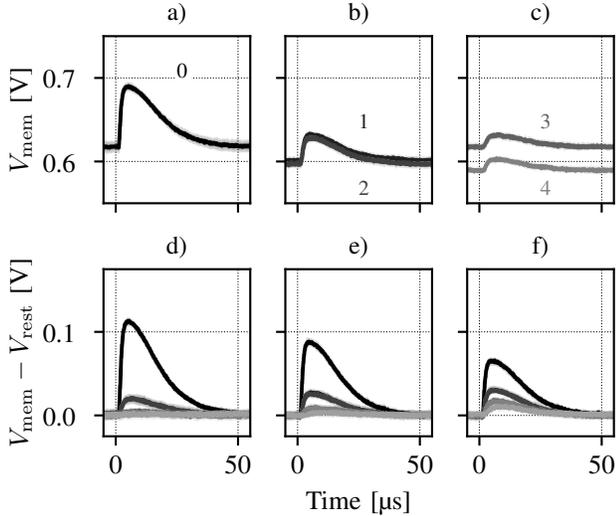

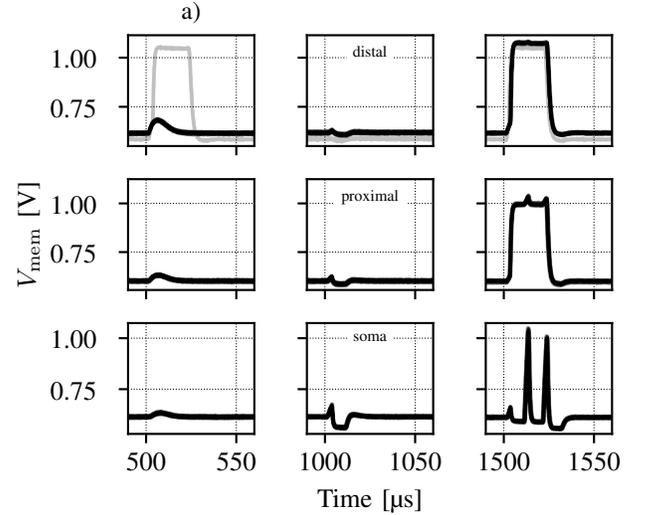

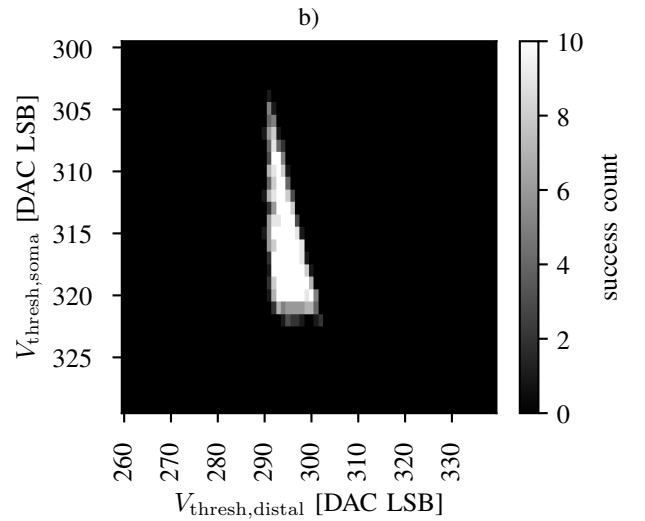

Fig. 9: Five neighboring neuron circuits are configured to form a multi-compartment neuron. a)–c) The compartments 0–6 are interconnected in a chain. Each even compartment connects via $R_{ic}$ to the shared line, whereas each odd compartment connects to the shared line and, using $S_{ic}$, to its right neighbor. Compartment 0 is stimulated via the synaptic input at time 0. The attenuated signal can be observed in consecutive compartments. b) Because two neighboring compartments, starting from compartment 1, are short-circuited, the same response is recorded over two output buffers. c) Due to different mismatch in the readout chain, the recorded signals show a varying degree of offset (see [31], [32] for offset characterization.) d)–f) Membrane voltage for all compartments in the chain after subtracting the baseline resting voltage $V_{\text{rest}}$. The current bias which controls the inter-compartment resistance is varied: d) $I_{\text{Ric}} = 50\,\text{LSB}$ e) $I_{Ric} = 150\,\text{LSB}$ f) $I_{\text{Ric}} = 1000\,\text{LSB}$. Each line shows one standard deviation acquired from ten consecutive measurements.

## VI. DISCUSSION

This paper summarized the AdEx and multi-compartment extensions developed to enhance the base 65 nm CMOS LIF neuron for the HICANN-DLS prototype chip. For point-neuron model enhancement, we integrated adaptation and exponential circuit to the modular LIF neuron architecture. The AdEx enhancement let us qualitatively reproduce exponential spikes, and diverse spiking and bursting regimes. Using a floating tunable resistor we can tune very long adaptation time constants. Further, by utilizing controlled inter-compartmental resistance and transmission-gate switches, we extend the AdEx emulation to multiple compartments. The conductance-based reset and

Fig. 10: Multi-compartment configuration. a) Membrane voltage of individual compartments in an interconnected neuron. Top row: distal compartment, center row: proximal compartment, bottom row: soma compartment. Left column: A single input spike is sent to the distal compartment. Center column: A single input spike is sent to the soma compartment, which evokes a spike event. Right column: Both stimuli are combined. The stimulus magnitude is unchanged. Each panel shows 20 superimposed voltage recordings. Because at most two compartments can be recorded simultaneously, the data shown in different rows stem from consecutive experiments. Thus, data from 60 experiments are shown in total. b) Parameter sweep of the firing threshold in the soma and distal compartments. The color indicates the number of successful repetitions out of ten trials. The success is evaluated using the number of spikes in the soma and distal compartments (see text).

| Neuron model | multi-comp. AdEx |
|---|---|
| No. of neurons | 32 |
| Area of single neuron | 11.76 × 286 µm² |
| Voltage supply | 2.5/1.2 V |
| Process | 65-nm CMOS |
| Speed-up (acceleration) factor | ×1000 |
| Shared parameters[1] | 3 current biases |
| Dedicated parameters[1] (per neuron) | 20 (14 I-bias, 6 V-bias) |
| Digital Configuration | 40-bit SRAM |
| Membrane capacitor (max.)[2] | 2.36 pF (6-bit config.) |
| Input synaptic event (max.)[3] | 10 µA, 4 ns pulses |

[1] available from on-chip tunable capacitive memory cells [39], [51]
[2] binary weighted with 37 fF unit capacitance; extendable to 4.36 pF if adaptation is not used
[3] amplitude and length of each current pulse emitted by the synapse circuit

TABLE I: A summary of neuron array specifications.

|  | TrueNorth [52] | Neurogrid [53] | This Work |
|---|---|---|---|
| CMOS tech. [nm] | 28 | 180 | 65 |
| Architecture | digital | analog subth. | analog |
| Speed-up[a] | ×1 | ×1 | ×1000 |
| Neuron Model | augmented LIF | QIF | AdEx |
| Dedicated Parameters | Yes | No | Yes |
| Biophysical Dynamics[b] | No | Yes | Yes |
| Neuron Area [µm²] | 2900[c] | 1800 | 3372 |

[a] compared to biological real-time
[b] tunable $\tau_{refr}$, $\tau_{mem}$, $\tau_{syn}$
[c] multiplexed 256 times per time step

TABLE II: An overview of neuron model specifications in large-scale neuromorphic architectures.

digital pulse control let us emulate the broad spikes (plateau potentials) known from apical and thin dendrites. We have demonstrated the functionality of emulated passive dendrites in the circuit by connecting multiple adjacent compartments. Furthermore, we have configured two adjacent neuronal compartments of an on-chip neuron as proximal and distal dendrites to demonstrate active coincidence detection that leads to plateau potentials in proximal dendrites and a burst of sodium-like spikes in the somatic compartment.

This prototype chip demonstrates a scaled-down AdEx multi-compartment neuron accelerated to 1000 times compared to biological real-time. Each neuron occupies 11.76 × 286 µm², including the digital circuits for pulse control, AdEx multi-compartment extensions and dedicated SRAM. This neuron circuit together with number of input synapses and analog bias memory is scheduled to scale up for a larger chip that will feature 512 neurons with increased silicon area. The achieved neuron specifications are listed in Table I.

The neuron circuit reduces the effect of device mismatch and resulting input offsets after calibration (see [32] for calibration of LIF circuits). The variation in adaptation time constant among individual neurons is to be calibrated for, although it does not preclude successful network operation. During the design phase, the possibility of calibration has been ensured for the circuit netlist using device Monte Carlo models. The simulated energy per output spike expended by the neuron circuit is 200 pJ for a 300 Hz output firing rate. The energy per synaptic event depends on the realized network configuration, number of input synapses, etc., and for the first-generation 180 nm wafer-scale system varies between 100 pJ and 10 nJ [54].

A plethora of silicon neuron designs can be found in literature – from simple LIF neurons to two-dimensional models and those implementing dendritic properties in analog circuits [38], [53], [55]–[70] – as well as digital phenomenological implementations [71]–[75]. Here we restrict our comparison to neuron circuits implemented for large-scale neuromorphic architectures, and summarized in Table II. The TrueNorth neuron [52], [73] features an augmented LIF neuron model in a scaled 28 nm CMOS process targeted mainly for synthetic computation. The implementation therefore does not feature tunable biophysical dynamics such as synaptic time constant or refractory period duration. The neuron occupies 2900 µm² despite a scaled process node, but the possibility of time-multiplexing reduces the effective area by 256 times. In comparison, Neurogrid [53] implements biophysical dynamics like the BrainScaleS system. The neuron model is a two-compartment quadratic model implemented using subthreshold MOS dynamics in a 180 nm CMOS process. The subthreshold implementation typically reduces the occupied area. The system is tiled in *NeuroCores*, each containing 64k neurons that share the neuron parameters. Both TrueNorth and Neurogrid system implement real-time dynamics compared to the accelerated implementation of the BrainScaleS system and the current work. Our current multi-compartment AdEx neuron emulation with its tunable dynamics present a biologically plausible neuromorphic platform. The distributed memory architecture with dedicated analog parameter memory and local SRAM in every neuron makes the system highly flexible, re-configurable and amenable to calibration that evolves as a profoundly non-von Neumann implementation.

## VII. CONCLUSION

In this work, we presented circuit extensions to enhance the previously designed modular LIF neuron architecture to multi-compartment AdEx model. The architecture presented a re-configurable circuit for firing configurations ranging from simple LIF to multi-compartment and AdEx modes. It allowed us to reproduce exponential and adapting spike response as well as diverse spiking firing patterns. By integrating multi-compartment extensions, we have demonstrated properties of passive and active non-linear neuronal dendrites. Having presented the advance neuron model, we march on to integrate a scaled-up HICANN-DLS chip – the building block of our second-generation neuromorphic platform.

ACKNOWLEDGMENT

The authors would like to express their gratitude to Andreas Hartel, Christian Pehle, Korbinian Schreiber, Sebastian Billaudelle and David Stöckel for the teamwork during design, verification and measurement of the prototype chip. Further, the authors would like to thank Eric Müller, Sebastian Schmitt and Mihai Petrovici for systems support and theoretical discussions.

The authors would like to thank Stefan Scholze and Sebastian Höppner from TU-Dresden for the PLL macro cell [76] used in the prototype chip.

AUTHOR CONTRIBUTION

S.A.A. wrote the manuscript, designed and measured AdEx neuron circuit array. P.M. devised the pre tape-out calibration, performed experiments on structured neurons and wrote Sec. V-D & Sec. V-E. G.K. designed the digital circuits for neuron pulse control. L.K. verified firing patterns on the pre tape-out netlist. Y.S. provided measurements on adaptation time constant and software support. A.G. was the chip assembly responsible. J.S. was the overall system architect and designed multi-compartment/non-linear dendritic extensions. The conceptual advice was given by K.M.

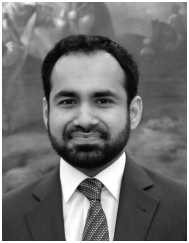

**Syed Ahmed Aamir** (S'11) received the M.Sc. degree from Linköping University, Sweden and Ph.D. degree from Karlsruhe Institute of Technology, Germany, both in electrical engineering in 2010 and 2018 respectively. He has worked as a Research Engineer at Linköping University and currently works as a Postdoctoral Researcher in Electronic Vision(s) group at Kirchhoff Institute for Physics, Heidelberg. His current research interests include low-power analog design, integrated mixed-signal/RF transceivers and biologically-inspired systems.

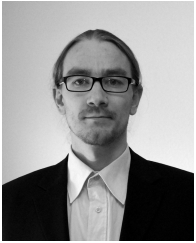

**Paul Müller** received the Dipl. Phys. degree from Heidelberg University, Heidelberg, in 2011. He was working as a research assistant at the Electronic Vision(s) group until 2013, and received his Ph.D. degree in Physics in 2017. His current research interest is the development, operation and application of mixed-signal neuromorphic devices.

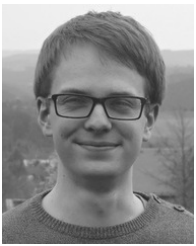

**Gerd Kiene** received the M.Sc. degree in Physics from Heidelberg University, Germany, in 2017. Currently he is working as a research assistant in the Electronic Vision(s) group at the Kirchhoff Institute for Physics developing analog readout and mixed-signal circuits for neuromorphic systems. His research interests are the development of novel computing systems as well as digital, analog and RF integrated circuits.

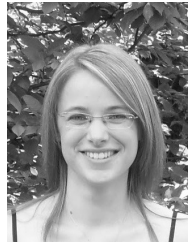

**Laura Kriener** received the M.Sc. degree in Physics from Heidelberg University, Germany, in 2017. Her Master's thesis in the Electron Vision(s) group at the Kirchhoff Institute for Physics focused on the pre-production characterization of single neuron dynamics on neuromorphic hardware. Her research interests are the development and application of neuromorphic hardware.

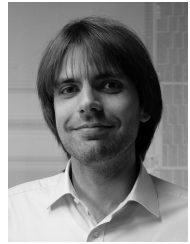

**Yannik Stradmann** received the B.Sc. degree in Physics from Heidelberg University, Germany, in 2016. His research as a Master's student within the Electronic Vision(s) group focuses on the development and verification of mixed-signal VLSI circuits for neuromorphic hardware.

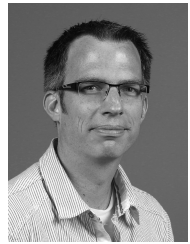

**Andreas Grübl** received the Dipl. Phys. and Ph.D. degrees from Heidelberg University, Heidelberg, Germany in 2003 and 2007, respectively. Currently, he is a Senior Postdoctoral Researcher in the Electronic Vision(s) group and leader of the Electronics Department of the Kirchhoff Institute for Physics at Heidelberg University. He has eight years of postdoctoral experience in designing and building complex microelectronics systems for brain-inspired information processing. His research focus is on new methods for the implementation of large mixed-signal SoCs.

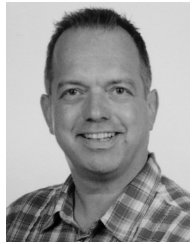

**Johannes Schemmel** (M'08) received the Ph.D. degree in physics from Heidelberg University, Heidelberg, Germany, in 1999. Currently, he is 'Akademischer Oberrat' in the Kirchhoff Institute of Physics, Heidelberg, where he is head of the ASIC laboratory and the Electronic Vision(s) group. His research interests are mixed-mode VLSI systems for information processing, especially the analog implementation of biologically realistic neural network models. He is the architect of the Spikey and BrainScaleS accelerated Neuromorphic hardware systems.

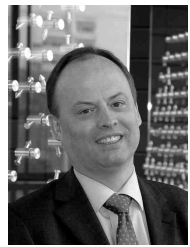

**Karlheinz Meier** (M'17) received the Ph.D. degree in physics from Hamburg University, Hamburg, Germany, in 1984. He is currently a Professor of Physics at Heidelberg University, Heidelberg, Germany, and Co-Founder of the Kirchhoff-Institut and the Heidelberg ASIC Laboratory in Heidelberg. His research interests include the application of microelectronics in particle physics, electronic realizations of brain circuits, and principles of information processing in spiking neural networks.